\documentclass[twocolumn]{aastex63}

\def\kms{km~s$^{-1}$}
\def\h2o{H$_2$O}
\def\nh3{NH$_3$}
\def\vlsr{$V_{\mbox{\scriptsize LSR}}$\ }

\def\hii{H{\rm II\ }}
\def\ngc{NGC\,6334I}

\def\13co{$^{13}$CO}
\def\c18o{C$^{18}$O}
\def\ch{CH$_3$OH}

\usepackage{natbib}
\usepackage{xcolor}
\usepackage[normalem]{ulem}
\usepackage{soul}
\usepackage{longtable}
\usepackage{amsmath}

\received{October 30, 2020}
\revised{November 10, 2020}
\accepted{17 December 2020}
\submitjournal{ApJ}

\shorttitle{NGC\,6334\,I-MM1: Spatio-kinematics of Water Masers during a Contemporaneous Flare Event}
\shortauthors{Chibueze et al.}

\begin{document}

\title{The Extraordinary Outburst in the Massive Protostellar System NGC\,6334\,I-MM1: Spatio-kinematics of Water Masers during a Contemporaneous Flare Event}

\correspondingauthor{James O. Chibueze}
\email{james.chibueze@nwu.ac.za}

\author[0000-0002-9875-7436]{James O. Chibueze}
\affiliation{Centre for Space Research, Potchefstroom campus, North-West University,
Potchefstroom 2531, South Africa}
\affiliation{Department of Physics and Astronomy, Faculty of Physical Sciences,  University of Nigeria, \\Carver Building, 1 University Road,
Nsukka 410001, Nigeria}
\author[0000-0002-1505-2511]{Gordon C. MacLeod}
\affiliation{Hartebeesthoek Radio Astronomy Observatory, PO Box 443, Krugersdorp 1741, South Africa.}
\affiliation{The University of Western Ontario, 1151 Richmond Street, London, ON N6A 3K7, Canada.}

\author[0000-0001-9572-2425]{Jakobus M. Vorster}
\affiliation{Centre for Space Research, Potchefstroom campus, North-West University,
Potchefstroom 2531, South Africa}

\author[0000-0003-1659-095X]{Tomoya Hirota}
\affiliation{National Astronomical Observatory of Japan,
National Institutes of Natural Sciences,
2-21-1 Osawa, Mitaka, Tokyo 181-8588, Japan.}
\affiliation{Department of
Astronomical Sciences, SOKENDAI (The Graduate University for Advanced
Studies), Osawa 2-21-1, Mitaka-shi, Tokyo 181-8588, Japan}

\author[0000-0002-6558-7653]{Crystal L. Brogan}
\affiliation{NRAO, 520 Edgemont Rd, Charlottesville, VA, 22903, USA.}

\author[0000-0001-6492-0090]{Todd R. Hunter}
\affiliation{NRAO, 520 Edgemont Rd, Charlottesville, VA, 22903, USA.}

\author[0000-0001-8940-4228]{Ruby van Rooyen}
\affiliation{South African Radio Astronomy Observatory, The Park, Park Road, Pinelands, 2 Fir Street, Black River Park, Observatory, 7925, South Africa}



\begin{abstract}

Following an eruptive accretion event in \ngc-MM1, flares in the various maser species, including water masers, were triggered. We report the observed relative proper motion of the highly variable water masers associated with the massive star-forming region, \ngc. High velocity H$_2$O maser proper motions were detected in 5 maser clusters, CM2-W2 (bow-shock structure), MM1-W1, MM1-W3, UCHII-W1 and UCHII-W3. The overall average of the derived relative proper motion is 85 \kms. This mean proper motion is in agreement with the previous results from VLA multi-epoch observations. Our position and velocity variance and co-variance matrix analyses of the maser proper motions show its major axis to have a position angle of $-79.4^\circ$, cutting through the dust cavity around MM1B and aligned in the northwest-southeast direction. We interpret this as the axis of the jet driving the CM2 shock and the maser motion. The complicated proper motions in MM1-W1 can be explained by the combined influence  of  the  MM1 northeast-southwest bipolar outflow,  CS(6-5)  north-south collimated bipolar outflow, and the radio jet. The relative proper motions of the H$_2$O masers in UCHII-W1 are likely not driven by the jets of MM1B protostar but by MM3-UCHII. Overall, the post-accretion burst relative proper motions of the H$_2$O masers trace shocks of jet motion.

\end{abstract}



\section{Introduction} \label{sec:into}
Accretion in young massive protostars is a complex phenomenon. Irregular and fragmented accretion disks lead to episodic accretion with long periods of relatively slow accretion rate, and short periods of high mass gain \citep{2017MNRAS.464L..90M}. High accretion rates (also called accretion bursts) have been directly observed in massive protostars of S255IR NIRS 3 \citep{2017NatPh..13..276C},
NGC6334I-MM1 \citep{2017ApJ...837L..29H} and G358.93-0.03-MM1 \citep{2020NatAs...4..506B}. The high accretion rate heats up the protostellar disk, which in turn can dramatically increase thermal radiation by the surrounding dust. Among the consequences of accretion events include enhancement of existing spectral line emission \citep{2018ApJ...854..170H,2018ApJ...866...87B,2020NatAs...4..506B} and the excitation of new maser lines \citep{2019ApJ...881L..39B,2019MNRAS.489.3981M,2020NatAs.tmp..144C,2020MNRAS.494L..59V,2020ApJ...890L..22C}. Accretion bursts can also significantly alter the chemical makeup of the protostellar disk for a short time, as observed in low mass protostars \citep{2015A&A...577A.102V}. 
\par
In general, astrophysical masers provide clues into the physical conditions and the kinematics of protostellar systems. Maser flares have been found to accompany accretion bursts initially identified in many observations: in 6.7\,GHz \ch{} masers near S255IR \citep{2017A&A...600L...8M,Szymczak2018} and many maser species from NGC6334I \citep{2018MNRAS.478.1077M}. Long term single-dish monitoring observations of variable masers can  provide an excellent mechanism to identify the onset of an accretion burst. The early identification of an accretion burst via maser monitoring was detected in G358.93-0.03 by \citet{Sugiyama19}. This exciting result led to the detection of a wide range of methanol maser transitions including several not even predicted to exist \citep{2019MNRAS.489.3981M,2019ApJ...881L..39B,2019ApJ...876L..25B}. Monitoring of
19.967\,GHz \ch{} masers towards G358.93-0.03 are presented in \citet{2020MNRAS.494L..59V} and for other transitions, in MacLeod et al. (in preparation) and Yonekura et al. (in preparation).

High resolution multi-epoch Very Long Baseline Interferometry (VLBI) studies of maser proper motion measurements provide useful insights into the gas spatio-kinematics of protostellar disks, outflows and shocks \citep{2011A&A...526A..66M, 2012ApJ...748..146C, 2014MNRAS.442..148T}.
\par  
NGC\,6334\,I, located at a parallax distance of $1.30 \pm 0.09$ kpc \citep{2014ApJ...784..114C,2014ApJ...783..130R,2014A&A...566A..17W}, is a massive star forming region containing a massive protostar that has recently undergone an accretion burst. Millimeter and sub-millimeter observations using the Submillimeter Array (SMA) first identified four compact sources in NGC 6334I (MM1-MM4). MM1 and MM2 were the brightest dust sources while MM3 coincided with the ultra-compact HII region NGC 6334 F \citep{2006ApJ...649..888H}. Five more continuum sources (MM5-MM9) were later identified using the Atacama Large Millimeter/submillimeter Array (ALMA) and MM1 was also resolved into six continuum components (A-F) at 1.3 mm \citep{2016ApJ...832..187B}. Properties of the individual components were modelled, with several having high dust and brightness temperatures $T_{dust} > 300 \,K, T_{brightness} > 200\,K$. 
Continuum emission associated with MM1, MM3-UCHII and CM2 (located north of MM1)
was detected in the 5\,cm observations of \ngc\/ made  using the Karl G. Jansky Very Large Array (VLA)  \citep{2016ApJ...832..187B}. Comparison of the ALMA images with earlier SMA images revealed that the luminosity of MM1 increased by a factor of $l_{inc} = 70 \pm 20$ \citep{2017ApJ...837L..29H} between 2008 and 2015, with the centroid of the increase aligned with protostar MM1B. In a parallel discovery, MM1F, MM1G, MM1C and MM3-UCHII underwent their first observed activation of 6.7 GHz masers \citep{2018ApJ...854..170H}, contemporaneous with flaring of nine other maser transitions beginning in January 2015 \citep{2018MNRAS.478.1077M}. Followup 22 GHz H$_2$O\/ maser emission measurements using VLA identified flaring of water masers in a bow shock shape in CM2 \citep{2018ApJ...866...87B}. In contrast, the H$_2$O\/ masers previously seen surrounding MM1B were also damped significantly, likely due to increased dust temperatures. 

Single-dish observations in various CO and CS transitions have consistently shown a northeast-southwest (NE-SW) outflow at large scales ($\sim$0.5 pc) whose origin is centered on MM1 or MM2 \citep{1990A&A...239..276B,2000MNRAS.316..152M,2006A&A...454L..83L,2011ApJ...743L..25Q}. Later interferometric imaging with ALMA of CS(6-5) resolved the central part of the outflow, confirming MM1 as the primary origin, and revealing a north-south (N-S) outflow centered on MM1B and a blue-shifted northwest (NW) lobe \citep{2018ApJ...866...87B}.
Subsequent imaging of CS(18-17) and HDO in ALMA Band 10 \citep{McGuire2018} demonstrated excellent spatial alignment between the warm thermal gas tracing the compact outflow and the 22 GHz H$_2$O masers embedded in it.

\par
In this paper, we present multi-epoch VLBI measurements of the 22 GHz H$_2$O\/ maser emission using a combination of Korean VLBI Network (KVN) and VLBI Exploration of Radio Astronomy (VERA) during the first year of the accretion burst event. We derive the relative proper motion measurements for these H$_2$O\/ masers in order to probe the kinematics of the gas surrounding MM1 and MM3-UCHII in this region of active star formation.

\section{Observations and data reduction} \label{sec:obs} 
\subsection{Single-dish monitoring observations}
The ongoing 22.2\,GHz (1.3\,cm) water maser observations reported here were made using the 26m telescope of Hartebeesthoek Radio Astronomy Observatory (HartRAO). The single-dish results reported in this paper covers observations taken between 7 May 2013 and 2 December 2020. The coordinates that the telescope pointed to were ($\alpha,\delta$)=(17$^h$20$^m$53$^{s}$.4, $-$35$^o$47\arcmin01\farcs5). The beam width for this receiver is 2.2\farcm{} Pointing observations were made for each epoch. These observations were also corrected for atmospheric absorption. Because of the large velocity extent position switching was employed. The rest frequency of the receiver was set to 22.235120\,GHz. The receiver system consisted of left (LCP) and right circularly polarised (RCP) feeds. Dual polarization spectra were obtained using a 1024-channel (per polarisation) spectrometer. The receiver is cryogenically cooled. Each polarisation is calibrated independently relative to Hydra A, 3C123, and Jupiter, assuming the flux scale of Ott et al. (1994). The band width used was 8\,MHz providing a velocity resolution of 0.105\,\kms\/ and a total velocity extent of 107.9\,\kms. Typical sensitivities achieved per observation were 2.3 to 2.9\,Jy. Typically, observations were made every 10 to 15d. However, the cadence of observations varied depending on the availability of the telescope, and the weather conditions. At times observations were done daily, but there are also observations separated by weeks. 

\subsection{KaVA observations}
H$_2$O masers in \ngc~was observed with KVN and VERA Array (KaVA) in 3 epochs taken on 21 November 2015 (2015.89), 15 December 2015 (2015.95) and 4 January 2016 (2016.01), respectively. The position of the phase tracking center of the \ngc~was ($\alpha,\delta$)=(17$^h$20$^m$53$^s$.377, -35$^o$46\arcmin55\arcsec.808) in the J2000.0 epoch. NRAO530 was used as the band-pass calibrator while 

The total bandwidth was 256 MHz (16 MHz $\times$ 16 IFs) and data were recorded for the left-hand circular polarization at a 1 Gbps sampling rate. We analyzed only the one 16 MHz IF channel that contained the H$_2$O\,6$_{16}$--5$_{23}$ transition. The spectral resolution is 15.625 kHz ($\sim$0.21 \,\kms) for the H$_2$O\/ maser line. The correlation process was carried out at the Korean-Japan Correlation Center, Daejeon, Korea (KJCC: Lee et al. 2015).

The data calibration was carried out using the Astronomical Image Processing System (AIPS) developed by National Radio Astronomy Observatory (NRAO) (van Moorsel et al. 1996). First, the amplitude was calibrated by using AIPS task APCAL using system temperature and measured antenna gains. Next, delays and phase offsets were removed by running AIPS task FRING using NRAO530. Bandpass response was also calibrated using NRAO530. The 3.9 \,\kms\/ velocity component of the masers was used as a reference maser component in \ngc. Imaging and CLEAN (deconvolution) were performed using the AIPS task IMAGR. The SAD task was employed for the Gaussian fitting for extraction of the peak intensities and offset positions of the maser spots.
A maser ‘spot’ refers to an individual maser emission peak in a spectral channel while a maser ‘feature’ denotes a group of maser spots considered to exist within the same maser cloudlet and located physically close to each other. The synthesized beams for the first, second, and third epochs was 2.48\,mas $\times$ 0.97\,mas (position angle, PA=-0.38$^\circ$), 2.66\,mas $\times$ 1.01\,mas (PA=2.82$^\circ$) and 2.76\,mas $\times$ 1.06\,mas (PA=10.64$^\circ$), respectively.

Maser features, defined as clusters of masers spots having a position defined by the position of the brightest peak, were carefully identified in each epoch. Maser distributions in MM1-W1 and UCHII-W1 varied significantly from epoch to epoch, and this complexity may have affected, in small measures, the derived proper motions. Our  single-dish results support the complex structures of these masers (see Section \ref{sec:single-dish}). The proper motions $\mu_{x}$ in R.A., and $\mu_{y}$ in Dec. were calculated using the displacement ($\Delta \alpha $cos$ \delta, \Delta \delta$) of the maser feature over adjacent epochs. For features detected in all three epochs, the average of the proper motion between epochs 1 \& 2, and epochs 2 \& 3 were taken.

Fringe-rate mapping was used to derive the absolute position of the reference maser spot and then compared to the closest epoch of the Very Large Array (VLA) H$_2$O maser map to obtain the absolute positions of the maser spots/features. The positional accuracy of the masers are within 1.0 mas in R.A. and 3.5 mas in Declination. To register the relative positions of the maser features in the 3 epoch, we use the position of a bright maser spot in UCHII-W1 region as a reference. The derived relative proper motions are marginally affected by the intrinsic motion of the reference maser spot. The overall uncertainty in our derived relative proper motions due to the motion of the reference maser spot is $<$\,10\%. This is obtained from the group motion of all maser features around the reference maser spot. It should be noted that all proper motions reported in this work are relative proper motions.

\section{Results} 
\label{sec:results}
\subsection{Structures in the H\texorpdfstring{$_2$}OO maser dynamic spectra}
\label{sec:single-dish}
\begin{figure*}[h]
\gridline{\fig{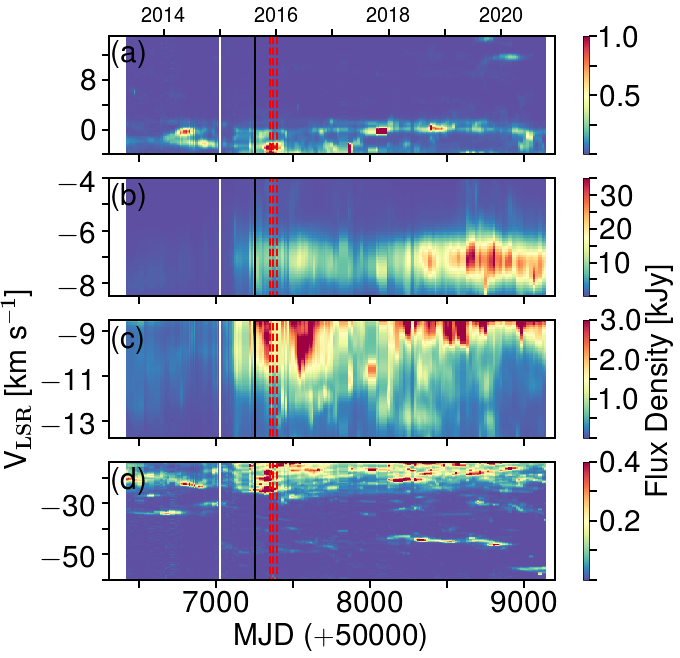}{0.45\textwidth}{(A)}
          \fig{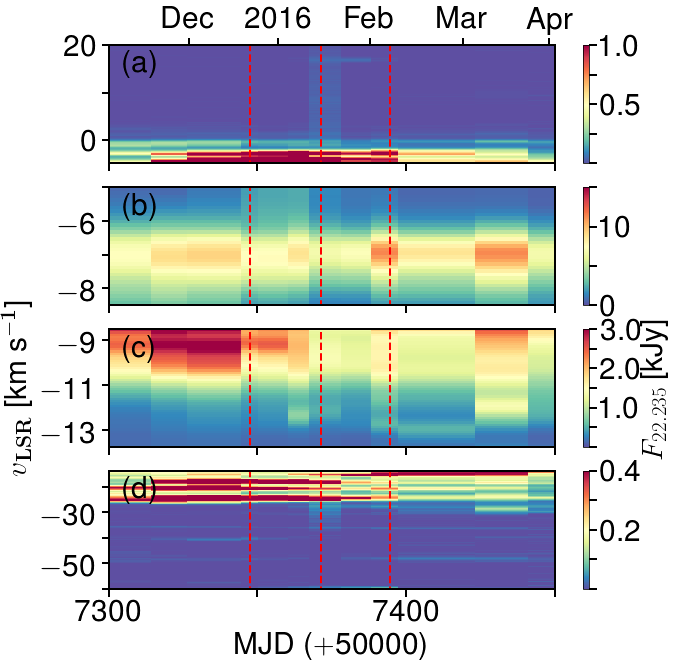}{0.45\textwidth}{(B)}
          }
\caption{(A) Dynamic spectra of the water masers associated with \ngc for the velocity extent (a) $-$4 to $+$15 km/s, (b) $-$8.5 to $-$4 km/s, (c) $-$13.5 to $-$9.5 km/s, and (d) $-$60 to $-$14 km/s. The white solid lines indicates 01 January 2020 (MJD 57023.5) marks the onset of the 6.7 GHz \ch maser burst and the black solid
lines indicates 15 August 2020 (MJD 57249.5), which marks the first maximum of the bursting masers presented in MacLeod et al. (2018). The dashed red lines mark the dates of each epoch of VLBI observations reported here. (B) Zoom-in image of A showing a close-up view of the dynamic spectra around the dates of our VLBI observations. The zoom-in shows the maser features varied in their intensities but were present in all 3 VLBI epochs.}
\label{fig:dyn_spec}
\end{figure*}

The dynamic spectra of the long-term monitoring of H$_2$O masers is shown in Fig.\ref{fig:dyn_spec} (A and B). The image provides an interesting metric demonstrating the longevity of emission in a given velocity extent. We note that a subset of this data was presented in \citet{2018MNRAS.478.1077M}. The water emission in $-$14$\leq $ V$_{LSR} \leq -$4 \kms\/ suffers significant line blending making it impossible to disentangle maser features and structure in this single-dish data. In panels (a) to (d) more independent masers are visible. In these, and during the MJD extent between the first and last epoch of VLBI observations, little velocity drift appears present in most. Possible velocity drifts may be present in this MJD extent for emission in $-$45$ \leq $ V$_{LSR} \leq -$35 \kms. This may be the result of multiple masers varying independently. Still the continuity of maser emission during this MJD extent lends comfort to the study of proper motion below.

Between the onset of the 6.7 GHz \ch maser burst and the first maximum of the burst (white and black lines in Fig.\ref{fig:dyn_spec} A, respectively), H$_2$O masers in the region are mostly destroyed or heavily suppressed. However, Fig. \ref{fig:dyn_spec} B shows that most of the maser features, though varied in the flux densities, survived through the epochs of our VLBI observations.

\begin{figure*}
\centering
\includegraphics[width = 0.62\textwidth]{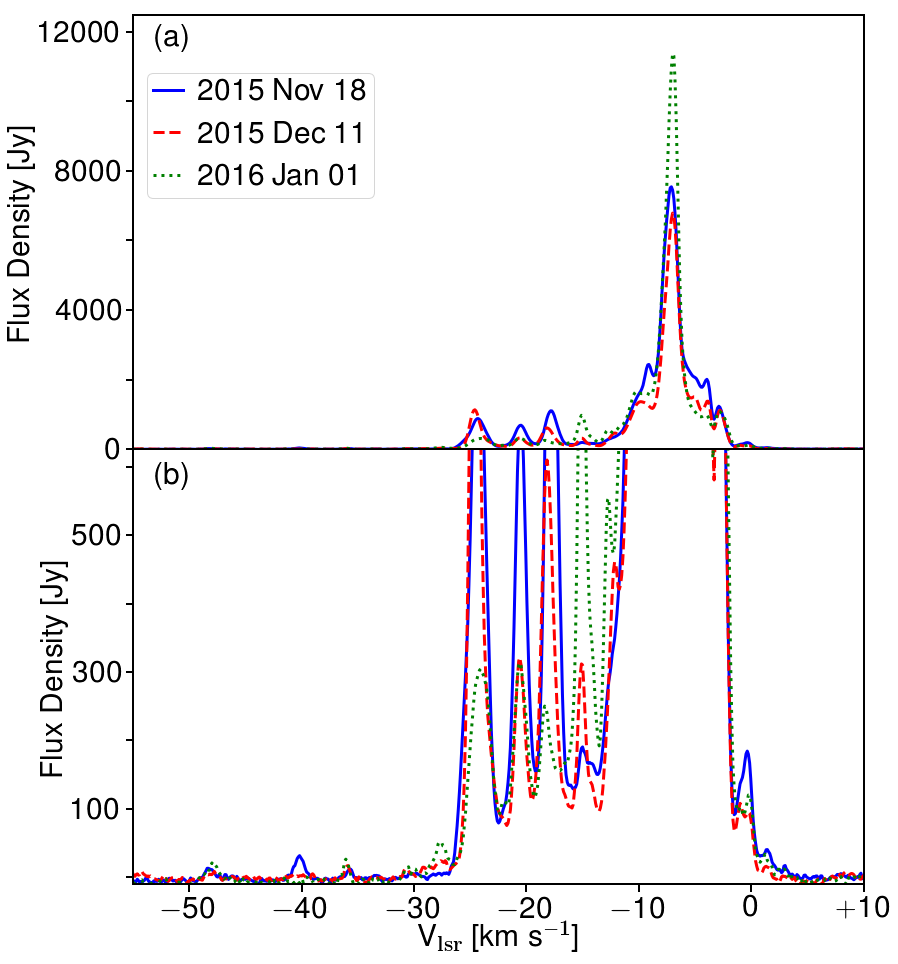}
\caption{Single-dish H$_2$O maser spectra of \ngc taken with HartRAO 26\,m closest (within $\pm$ 4 days) to each of the 3 epochs of our KaVA observations. (a) shows the full spectra and (b) shows the zoom-in into the weaker maser features.}
\label{fig:3_spec}
\end{figure*}

Figure \ref{fig:3_spec} shows the single-dish (HartRAO 26\,m) spectra of the highly variable H$_2$O masers in \ngc\/ taken nearest the respective VLBI observations. Significant variations can be seen between 18 November, 2015 and 01 January, 2016. The most prominent feature of the spectra, $-$7 \kms, and a second,  $-$15 \kms, feature are brightening, the rest of the maser features are weakening.

\subsection{Proper motions of the H\texorpdfstring{$_2$}OO masers} \label{sec:propmotions}
We obtained the absolute position of the reference maser spot (used for the self calibration) in the first epoch. This position, with full consideration of the proper motion of the reference maser spot, was used for the registration of the maps in the 3 epochs. We traced 186 maser proper motions, divided into groups according to the nomenclature used by \citet{2018ApJ...866...87B}. 

\begin{figure*}
\centering
\includegraphics[width = 0.6\textwidth]{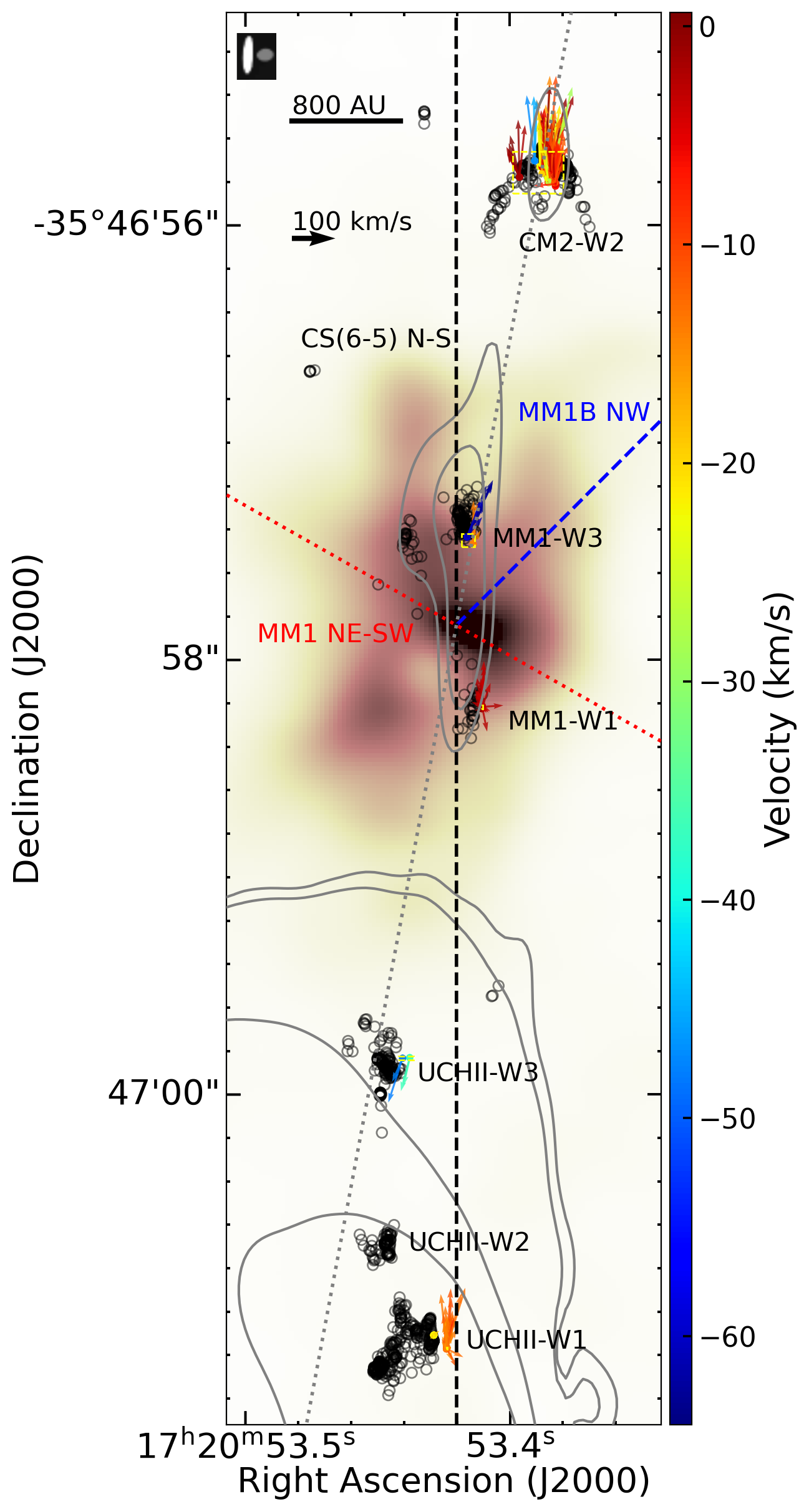}
\caption{H$_2$O maser proper motions derived from our KaVA observations overlaid on 2016.6 ALMA 1.3 mm continuum (brown scale) \citep{2017ApJ...837L..29H}. Grey contours are 2016.9 VLA 5 cm continuum observations with levels 0.022 $\times$ [4,9,260,600] mJy beam$^{-1}$ \citep{2018ApJ...854..170H}. H$_2$O maser regions are named according to the corresponding maser groups of \citet{2018ApJ...866...87B} from north to south (black labels). The blue dashed line shows the axes of the MM1B NW jet. The red dotted line shows the NE-SW wide angle outflow from MM1 and the black dashed line shows the outflow traced in CS(6-5). The black circles trace water masers measured by VLA in the 2017.8 epoch of \citet{2018ApJ...866...87B}. The grey dotted line shows the main velocity axes derived from the VVCM analysis (see Section \ref{subsec:VVCM analysis}).The linear scale and the transverse velocity scale is shown in the top-left corner. The radial velocity of the proper motions is indicated by the color scale. The synthesized beams are shown in the top left corner, where the white and grey ellipses are VLA and ALMA's beams respectively. The offsets (visible in zoom-ins) in the positions of the VLA 2017.8 maser features (black circles) could be due to error in the absolute position of our KaVA reference maser spot and/or the relative position uncertainty (19 mas in R.A and 66 mas in Declination) of the VLA observations.}
\label{fig:pm_all}
\end{figure*}

\begin{figure*}
\centering
\includegraphics[width = 0.85\textwidth]{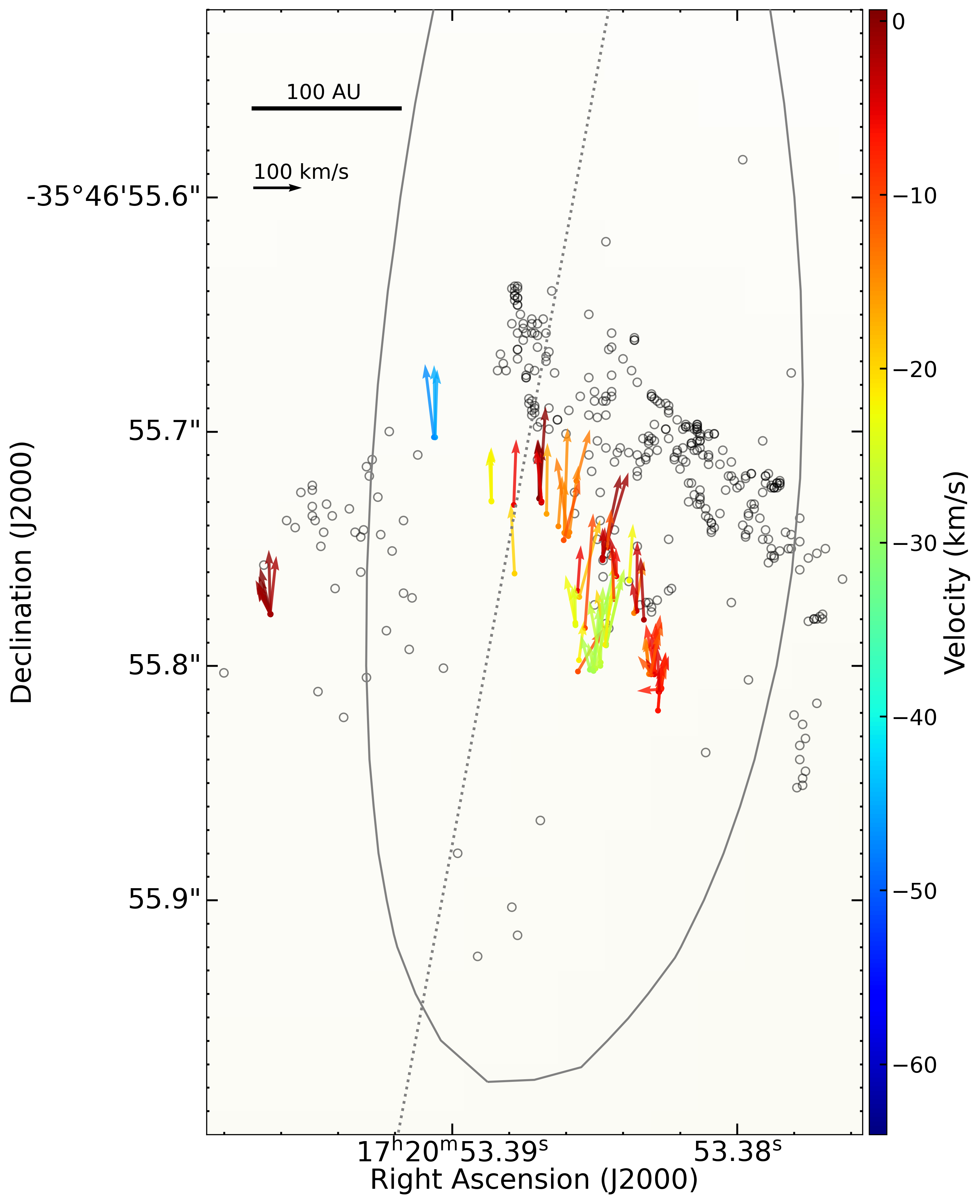}
\caption{Zoom-in of H$_2$O maser proper motions associated with the CM2-W2 region (See Figure \ref{fig:pm_all}). The \vlsr{} scale, contour lines, black circles and grey dotted line are the same as in Figure \ref{fig:pm_all}. Linear distance and velocity scale is shown in the top left. The seeming large offset in the positions of the VLA 2017.8 maser features (black circles) could be due to error in the absolute position of our KaVA reference maser spot and/or the relative position uncertainty (19 mas in R.A and 66 mas in Declination) of the VLA observations.}
\label{fig:CM2-W2}
\end{figure*}

\begin{figure*}
\centering
\includegraphics[width = 0.85\textwidth]{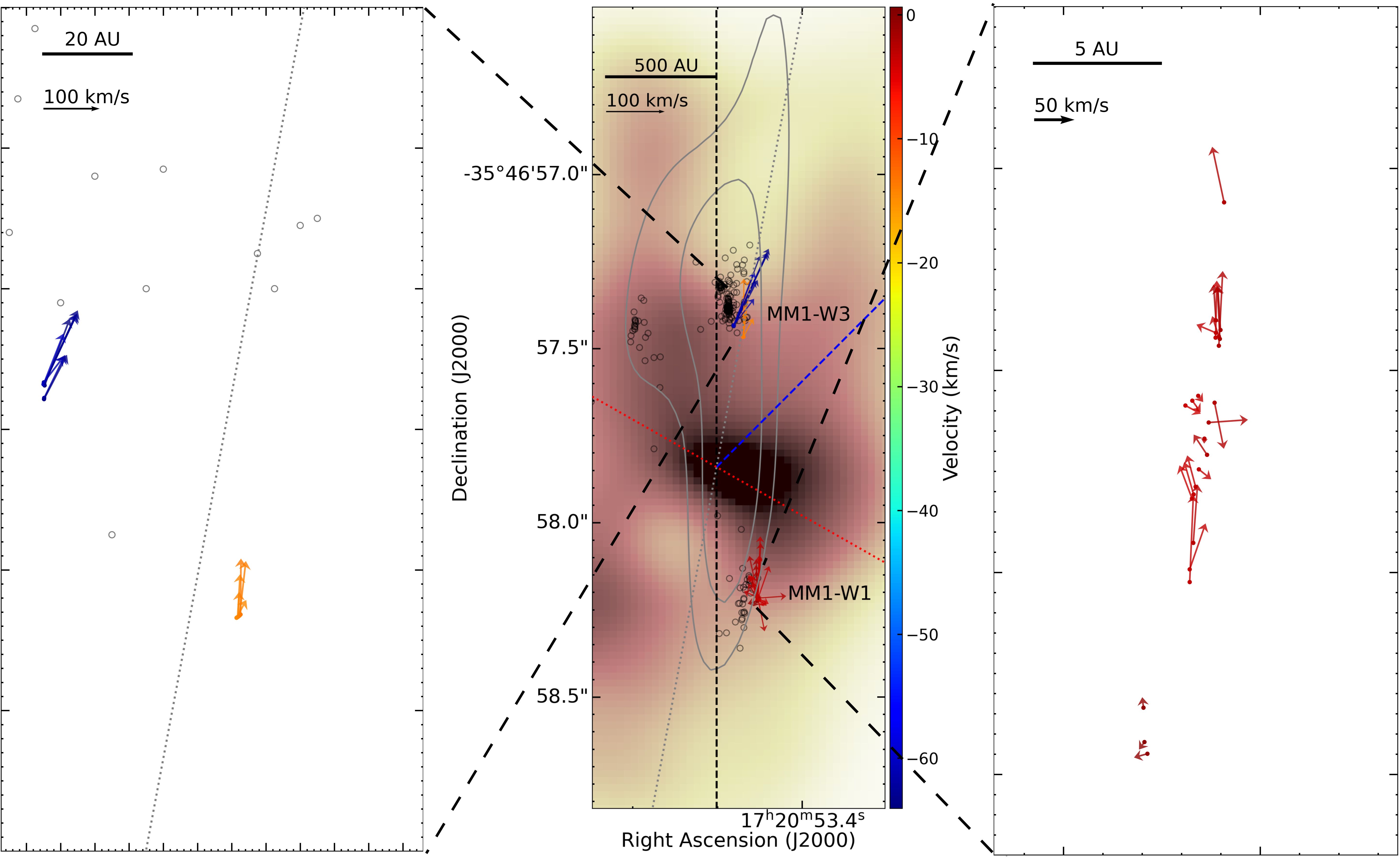}
\caption{Zoomed-in image of the MM1 region. With a high resolution image of the proper motions of the MM1-W1 region (right) and MM1-W3 (left). The \vlsr{} scale is shown by the color bar of the center image. Contour lines, black circles and grey dotted line are the same as in Figure \ref{fig:pm_all} for both images. Continuum was removed from the zoomed images for clarity.
Linear and velocity scales are shown in the top left corner of each image.}
\label{fig:MM1}
\end{figure*}

\begin{figure*}
\centering
\includegraphics[width = 0.85\textwidth]{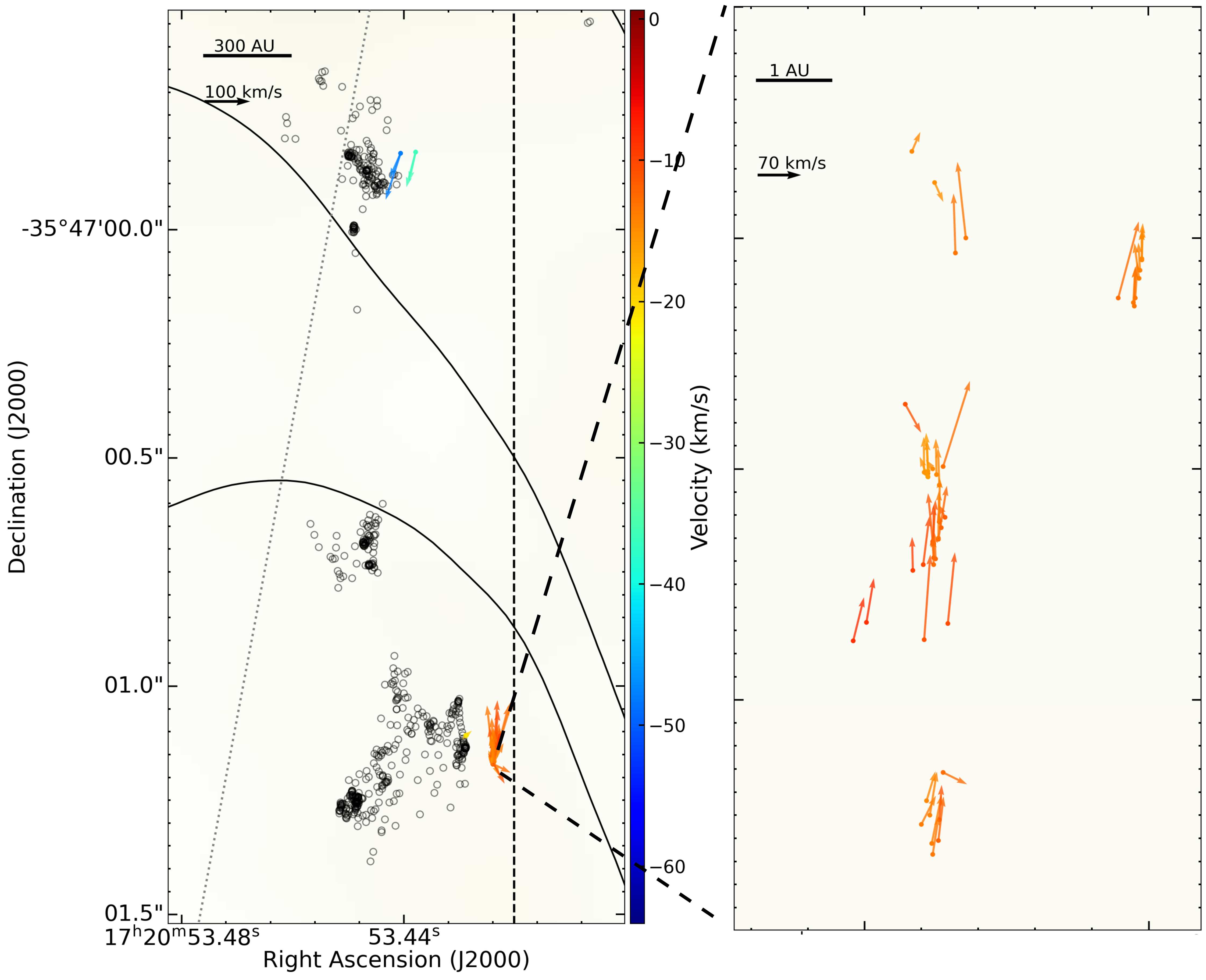}
\caption{Left: Zoomed in image of the UCHII region. Right: High resolution image of the proper motions of the UCHII-W1 region. The \vlsr{} scale for both images is shown on the colorbar of the left image. Contour lines, black circles and grey dotted line are the same as in Figure \ref{fig:pm_all} for both images. Linear distance and velocity scale is shown in the top left of both images.}
\label{fig:UCHII}
\end{figure*}

Figures \ref{fig:pm_all}, \ref{fig:CM2-W2}, \ref{fig:MM1}, and \ref{fig:UCHII} show the traced H$_2$O\/ maser proper motions overlaid on the ALMA 1.3\,mm dust continuum from a comparable epoch (2016.6, grey scale from \cite{2017ApJ...837L..29H}) and the VLA 5\,cm image (white contours from \cite{2018ApJ...854..170H}). The colored vectors (arrows) represent the H$_2$O\/ proper motions traced in the region. The length of each arrow indicates the magnitude of the proper motion and the direction of the arrow indicates the proper motion direction. Proper motions are measured with respect to the reference maser spot in UCHII-W1 located at $(\alpha,\delta)$ = $(\alpha,\delta)$ = (17$^h$20$^m$52$^s$.600, -35$^o$46\arcmin50\arcsec.508). The grey circles are water maser detection from \citet{2018ApJ...866...87B} for comparison. Figures \ref{fig:CM2-W2}, \ref{fig:MM1} and \ref{fig:UCHII} shows zoom-ins of the proper motions of different maser groups with colors of the \vlsr\/ of the masers (as in Figure \ref{fig:pm_all}).

We detected water maser proper motions in the regions CM2-W2, MM1-W1, MM1-W3, UCHII-W1 and UCHII-W3. The positions, proper motions, \vlsr{} and epochs of detection are shown in Table \ref{tab:pm_data}. A `+' indicates a detection in a specific epoch of a specific maser feature, while `-' signifies a non-detection. The majority
of all the proper motions (56.5\%) were traced using all 3 epochs, while the remaining 43.5\% were traced in 2 epochs. The overall mean of the 3D velocities of the masers is 85 \kms. For the rest of this section, average velocity refers to the magnitude of the average 3-dimensional velocity. In the following paragraphs, we describe the properties of each of the groups.

The northernmost region, CM2-W2, is $\sim$ 2750 AU from MM1B with 87 proper motions. Figure \ref{fig:CM2-W2} shows the spacial distribution, and proper motions of the H$_2$O masers in the region. There were 39 maser features detected in all three epochs. The proper motions have a spatial distribution comparable to a bow-shock shaped structure. Most of the proper motions point north, with an average velocity of 112\,\kms. The region also shows a drastic \vlsr{} gradient throughout the structure with $-46.98 <$ \vlsr{} $<$ 0.63 \kms. The proper motions detected spanned a linear size of $\sim$219 AU from east to west.

MM1-W1 is found just below MM1B ($\sim$ 510 AU) and is a more complicated region, with proper motions pointing in various directions. The maser spots showing a linear structure with a length of $\sim$ 18 AU. Figure \ref{fig:MM1} shows high resolution images of the spatial distribution and proper motions of water masers in MM1-W1 and MM1-W3. We detected 25 proper motions, 14 traced in three epochs. The average velocity of the region is 43\,\kms\/ and $-0.21 < $ \vlsr $< -3.8$ \kms. The region shows great variation in proper motion direction and magnitude over a relatively small region although there is not a large \vlsr\/ gradient. The region contains a number of high velocity proper motions pointing northward with an average velocity of 54\,\kms. The complexity of the observed proper motions can be attributed to the combined influence of the MM1 northeast-southwest, CS(6-5) north-source bipolar outflows, and the radio jet. The relative error in proper motion of this region is only $\sim$20\% for most of the constituent proper motions, indicating that the proper motions do reflect multiple influences on the motion of the masing cloudlets in the region.

MM1-W3 is the maser group just north of MM1B ($\sim$ 510 AU). We detected 13 proper motions with an average velocity of 106\,\kms. The region has two distinct associations. The north-eastern association consists of 8 proper motions, 5 are traced in all three epochs. The proper motions point north-west with an average of 126\,\kms\/ and a radial velocity \vlsr $\approx -62 $\kms. The second association points northward, with 4 of the 5 proper motions only being traced in two epochs. The average velocity of the association is 72\,\kms\/ and \vlsr $\approx$ 14 \kms. The linear separation between the two associations is $\sim$ 55 AU.
UCHII-W1 is about 4300 AU south of MM1B. We detected 48 proper motions, with 37 traced in three epochs. The region has an average velocity of 64\,\kms. There is a small radial velocity gradient with $-16.0 < $ \vlsr $< -8.2$ \kms. It should be noted that the maser spot distribution in this region was very complicated and the tracing of proper motions was difficult. Figure \ref{fig:UCHII} shows the spacial distribution and proper motions of masers in the MM3-UCHII region and a high resolution image of proper motions in UCHII-W1. Our results show a bulk motion to the north. 

UCHII-W3 is well south of MM1 ($\sim$ 2600 AU), corresponding to the edge of a jet traced by a CS (6-5) map from \citet{2018ApJ...866...87B}. We detected 4 proper motions with an average velocity of 89\,\kms\/ pointing to the south-east. Two maser associations are resolved $\sim$ 43 AU apart, the eastern association has an average velocity of 96 \kms\/ and \vlsr $\approx -48$ \kms. The western region has an average velocity of 81\,\kms\/ and \vlsr\/ $\approx -36$\,\kms.

\section{Discussion} \label{sec:discussions}

\subsection{VVCM analysis}
\label{subsec:VVCM analysis}

In order to characterize the proper motions of the outflow, we used the position variance-covariance matrix and velocity variance-covariance matrix (PVCM and VVCM) as described by \citet{1993ApJ...406L..75B,2000ApJ...533..893B} and \citet{2012ApJ...748..146C}.  These matrices provide a robust and objective means of extracting the position and kinematic essentials from maser proper motions. The PVCM and VVCM, $\sigma$, are constructed using:
\begin{equation}
\centering
    \sigma_{i,j} = \frac{1}{N-1}\sum^N_{n=1}(v_{i,n} - \bar{v_i})(v_{j,n} - \bar{v_j})
\end{equation}
with $i,j$ iterating over the spatial axes ($\alpha, \delta$ for the position variance-covariance matrix) and ($v_\alpha, v_\delta$,\vlsr{} for the velocity variance-covariance matrix), $n$ the $n$th of $N$ maser spots/proper motions ($N$=186). The bar indicates the average over all proper motions. The diagonal entries of the matrix $\sigma$ is the variance of the variable while the off-diagonal entries are the covariance of two variables.

The PVCM gives a 2\,$\times$\,2 matrix and using all the regions except UCHII-W1, the PVCM (in units of $10^{-6}$ arcsec$^2$) and its diagonalization was obtained to be:
\begin{equation}
    \begin{pmatrix}
  0.034 &  -0.163 \\
  -0.163 & 0.867
    \end{pmatrix} 
    \Rightarrow
\begin{pmatrix}
    0.003 & 0 \\
    0 & 0.897
\end{pmatrix}
\end{equation}

The corresponding 3\,$\times$\,3 VVCM matrix  and its diagonalization (in units of km$^2$ s$^{-2}$) is given by:
\begin{equation}
\begin{pmatrix}
   454.48 & -476.45 &  132.01 \\
    -476.45 & 3431.65 & -108.46 \\
    132.01 & -108.46 & 279.21
\end{pmatrix} 
\Rightarrow
\begin{pmatrix}
   3511.08 & 0 & 0\\
    0 & 452.48 & 0 \\
    0 & 0 & 201.79
\end{pmatrix}
\end{equation}
Table \ref{tab:PVCM/VVCM_results} shows the results of a PVCM and VVCM analyses. In Table \ref{tab:PVCM/VVCM_results}, $\psi_{max}$ indicates the largest eigenvalue of the PVCM/VVCM matrix, $\psi_{min}$ the smallest eigenvalue and $\psi_{mid}$ the middle-valued eigenvalue for the VVCM matrix. The large difference in the magnitudes of the eigenvalues of both position and velocity variance matrices demonstrates the presence of a distinct spatial and kinematic axis in the data. The major axis is defined by the eigenvector corresponding to the largest eigenvalue. The position angle is calculated by projecting the major axis onto the celestial sphere. The axis from the VVCM is plotted on Figure \ref{fig:pm_all} (and its zoom-ins) with a P.A. of -79.4$^\circ$ and passing through the position of MM1B from \citet{2016ApJ...832..187B}. The error in the position angle was calculated using a Monte-Carlo error of the velocity vectors. UCHII-W1 was not included as the direction of its motion does not seem to be influenced by the jet from MM1B. It should also be noted that including UCHII-W1 into the calculation makes only a marginal difference in the results ($\Delta$PA$_\text{max} \sim$ $ -2.25^\circ$, $\Delta \phi_\text{max} \sim $ $ 5.88^\circ $). The axis derived aligns very well with a bipolar outflow terminating at CM2-W2 and UCHII-W3. Assuming the bow shock in CM2 is symmetric, the inferred inclination angle for this outflow from matrix 2 is $\phi_\text{max} = -6.0^\circ \pm 0.6^\circ$ 

\subsection{Jet, cavity and shock structures in MM1}\label{sec:spex}

High proper motions of H$_2$O\/ near the path of the radio jet of Cepheus\,A-HW2 is attributed to the influence of the fast moving jet \citep{2011MNRAS.410..627T}. Typical proper motions of low velocity outflows and expanding ring/bubble structures are $\sim$10\,\kms \citep{2011MNRAS.410..627T,2012ApJ...748..146C,2014ApJ...784..114C}. With the mean H$_2$O\/ maser proper motion of 86\,\kms, the masing cloudlets in \ngc are driven by the jet in MM1.

Our VVCM analysis indicate the northwest-southeast axis of the jet driving the maser proper motions (at least of CM2-W2, MM1-W3, MM1-W1, UCHII-W3) as shown with dotted gray lines in Figure \ref{fig:pm_all}. Interestingly, this axis cut through dips in ALMA dust continuum, one in the northwest and the other in the southeast. We interpret these dips as cavities ploughed by the jets and this agrees with the suggested excavated outflow cavity by \citet{2018ApJ...866...87B}.

To test the possibility of precession in the jet motion, we compare the position angle of the VVCM results derived with all maser regions with those of the inner regions. About 10$^\circ$ difference is observed between the two position angles. This could be an indication of jet precession. MM1-W1 and MM1-W3 masers are closer to MM1 (driving source of the jet) and assuming a jet velocity of 150\,\kms, it will take 95 years for a jet launched by MM1 to reach the location of CM2. 

The synchrotron continuum point source CM2 \citep{2018ApJ...866...87B}, located north-west of the radio jet of MM1B, host the bright masers in the region and its nature has been discussed in \citet{2018ApJ...866...87B}. The observed proper motions of H$_2$O\/ masers in CM2 is similar to those reported in \citet{2016MNRAS.460..283B}. In a study of S255IR-SMA1 they reported a bow shock shape traced in H$_2$O masers with a velocity of $\sim$ 20 \kms{}, and a \vlsr{} gradient throughout the shock. They also reported three distinct ejections, with the most recent ejection being the shock traced in H$_2$O masers with a dynamical timescale $t_{dyn} \leq 130$ years. \citet{2017MNRAS.469.4788O} also reported a bow-shock structure for IRAS 20231+3440 traced with H$_2$O masers, with an average maser velocity of 14.26 \kms{}. These studies report bow-shock maser velocities significantly lower than we found in \ngc{}. Further studies into the driving mechanisms of the jets and outflows of \ngc{} and other sources are necessary to explain the bow-shock velocity discrepancies. This and the above mentioned studies (among others) indicate that a high velocity (V$_{ave} \geq$\,10 \kms{} H$_2$O maser proper motions in a bow-shock shape might be a common tracer for jets in massive protostars.

\subsection{Impact of MM3-UCHII on UCHII-W1 maser spatio-kinematics}

\citet{2018ApJ...866...87B} reported a bulk motion of 112$\pm$12\,\kms for the UCHII-W1 maser group using multi-epoch VLA observations between 2011 (pre-burst) and 2017 (post-burst). They suggested that the H$_2$O masers of the two 2017 epochs are possibly pumped by the beamed radiation from MM1B. The proper motions of UCHII-W1 point northward against the direction of the jet. This suggests that spatio-kinematics of the masing gas in this sub-region is not driven by the jet but by the MM3-UC\hii.

We investigated the possibility that the magnetic field reversal reported by \citet{2011MNRAS.414.1914C} and \citet{2018ApJ...854..170H} are responsible for the northward proper motion of UCHII-W1 H$_2$O masers. A reversal in magnetic field is reported in OH masers in UCHII-OH6 (located 0.35$\arcsec$ south-west of UCHII-W1) and UCHII-OH7  (located 0.7$\arcsec$ south of UCHII-W3) (see Figure 5 and Table 8 of \citet{2018ApJ...854..170H}).
 The reversed Zeeman splitted OH masers are $>$ 500 au from UCHII-W1 and rather closer to UCHII-W3 and W2 (see \cite{2018ApJ...866...87B}), therefore may not be responsible for the observed northward proper motions of UCHII-W1 masers. The UCHII-W1 proper motion is likely driven by MM3-UCHII or by the outflow of TPR-9 (with X-ray counterpart, CXOU 172053.21-354726.4) infrared star \citep{1996A&A...316..102T} as suggested by \cite{2018ApJ...866...87B}.

\section{Conclusions and summary}
We reported for the first time the spatio-kinematic of H$_2$O\/ masers in a massive star-forming region (\ngc) just after an accretion event. The proper motions of the H$_2$O\/ masers in CM2-W2, MM1-W3, MM1-W1, UCHII-W3 are mostly driven by the radio jet of MM1-B. However, some influence from the outflowing gas in MM1 NW-SE bipolar outflow and MM1B NW outflow cannot be completely excluded.

Our results suggested that the motion of the UCHII-W1 H$_2$O maser group is largely driven by the expansion of MM3-UC\hii. The significance of impact of the accretion event on the proper motions of the H$_2$O\/ maser, with special consideration of the destruction and re-excitation of the H$_2$O masers in the region,  will be presented in Vorster et al. (in prep.), which will compare the pre-burst and post-burst H$_2$O maser proper motions. The impact of a heat wave, such as the one reported in \citet{2020NatAs...4..506B}, will be explored with the pre-accretion burst VLBI H$_2$O maser data.

\acknowledgments

JOC acknowledges support from the Italian Ministry of Foreign Affairs and International Cooperation (MAECI Grant Number ZA18GR02) and the South African Department of Science and Technology’s National Research Foundation (DST-NRF Grant Number 113121) as part of the ISARP RADIOSKY2020 Joint Research Scheme. T. Hirota is financially supported by
the MEXT/JSPS KAKENHI Grant Number 17K05398. This paper makes use of the following ALMA data: ADS/JAO.ALMA\#2015.A.00022.T. ALMA is a partnership of ESO (representing its member states), NSF (USA) and NINS (Japan), together with NRC (Canada) and NSC and ASIAA (Taiwan) and KASI (Republic of Korea), in cooperation with the Republic of Chile. The Joint ALMA Observatory is operated by ESO, AUI/NRAO and NAOJ. The National Radio Astronomy Observatory is a facility of the National Science Foundation operated under agreement by the Associated Universities, Inc. This research made use of NASA's Astrophysics Data System Bibliographic Services. The Hartebeesthoek 26-m telescope is operated by the South African Radio Astronomy Observatory, which is a facility of the National Research Foundation, an agency of the Department of Science and Innovation. 


\startlongtable
\begin{deluxetable*}{cccccccccccc}
\tablecaption{Parameters of the H$_2$O\/ Maser Proper Motions}
\tablewidth{\textwidth}
\tablehead{
\colhead{ID$^\text{a}$} & \colhead{Region} & \multicolumn{2}{c}{Offset$^\text{b}$} & \multicolumn{4}{c}{Proper Motion} & \colhead{Radial Motion} & \multicolumn{3}{c}{Detections} \\
& &$\alpha$  & $\delta$  & $\mu_{x}$ & $\sigma$ $\mu_x$ & $\mu_y$ & $\sigma \mu_y$ &  \vlsr & 1 & 2 & 3 \\ 
\cline{3-4} \cline{5-8} \cline{10-12}
& & \multicolumn{2}{c}{(\arcsec)} & \multicolumn{4}{c}{(mas yr$^{-1}$)} & (km s$^{-1}$) & \multicolumn{3}{c}{} 
}
\startdata
$1$&CM2-W2&$-0.656$&$5.305$&$-2.394$&$0.168$&$21.709$&$0.168$&$-27.594$&$+$&$+$&$-$\\
$2$&CM2-W2&$-0.656$&$5.305$&$-3.711$&$0.168$&$24.368$&$0.168$&$-28.226$&$+$&$+$&$-$\\
$3$&CM2-W2&$-0.688$&$5.304$&$-0.980$&$2.200$&$12.964$&$0.413$&$-5.477$&$+$&$+$&$-$\\
$4$&CM2-W2&$-0.640$&$5.364$&$0.000$&$0.213$&$17.487$&$0.213$&$-14.535$&$+$&$+$&$-$\\
$5$&CM2-W2&$-0.640$&$5.364$&$-1.400$&$0.231$&$35.020$&$0.215$&$-14.533$&$+$&$+$&$-$\\
$6$&CM2-W2&$-0.627$&$5.379$&$-1.281$&$0.373$&$17.487$&$0.413$&$0.422$&$+$&$+$&$-$\\
$7$&CM2-W2&$-0.628$&$5.377$&$1.658$&$0.213$&$21.105$&$0.213$&$-2.316$&$+$&$+$&$-$\\
$8$&CM2-W2&$-0.628$&$5.378$&$1.960$&$0.213$&$19.145$&$0.213$&$-4.844$&$+$&$+$&$-$\\
$9$&CM2-W2&$-0.627$&$5.380$&$-2.384$&$1.643$&$30.351$&$3.193$&$-0.206$&$+$&$+$&$-$\\
$10$&CM2-W2&$-0.677$&$5.330$&$-4.179$&$0.168$&$17.412$&$0.168$&$-14.112$&$+$&$+$&$-$\\
$11$&CM2-W2&$-0.678$&$5.331$&$-0.281$&$0.168$&$23.235$&$0.168$&$-3.578$&$+$&$+$&$-$\\
$12$&CM2-W2&$-0.678$&$5.331$&$1.919$&$0.168$&$9.897$&$0.168$&$-3.789$&$+$&$+$&$-$\\
$13$&CM2-W2&$-0.646$&$5.326$&$0.183$&$0.259$&$13.204$&$0.259$&$-24.010$&$+$&$+$&$-$\\
$14$&CM2-W2&$-0.648$&$5.337$&$-8.619$&$0.493$&$25.858$&$2.556$&$-18.533$&$+$&$+$&$-$\\
$15$&CM2-W2&$-0.674$&$5.344$&$-1.540$&$0.168$&$18.956$&$0.168$&$-22.327$&$+$&$+$&$-$\\
$16$&CM2-W2&$-0.660$&$5.354$&$-7.336$&$0.259$&$27.692$&$0.259$&$-0.627$&$+$&$+$&$-$\\
$17$&CM2-W2&$-0.660$&$5.354$&$-10.637$&$0.259$&$28.425$&$0.259$&$-1.259$&$+$&$+$&$-$\\
$18$&CM2-W2&$-0.602$&$5.378$&$0.075$&$0.861$&$18.165$&$0.576$&$-21.698$&$-$&$+$&$+$\\
$19$&CM2-W2&$-0.602$&$5.378$&$0.301$&$0.337$&$17.487$&$0.282$&$-22.119$&$-$&$+$&$+$\\
$20$&CM2-W2&$-0.602$&$5.378$&$0.904$&$0.213$&$16.432$&$0.213$&$-22.330$&$-$&$+$&$+$\\
$21$&CM2-W2&$-0.690$&$5.288$&$-1.448$&$0.297$&$15.272$&$0.493$&$-6.738$&$-$&$+$&$+$\\
$22$&CM2-W2&$-0.662$&$5.317$&$4.000$&$0.168$&$18.241$&$0.168$&$-20.010$&$-$&$+$&$+$\\
$23$&CM2-W2&$-0.662$&$5.316$&$-7.476$&$0.168$&$25.572$&$0.168$&$-26.330$&$-$&$+$&$+$\\
$24$&CM2-W2&$-0.662$&$5.316$&$-6.602$&$0.168$&$23.152$&$0.168$&$-26.541$&$-$&$+$&$+$\\
$25$&CM2-W2&$-0.662$&$5.316$&$-0.603$&$0.213$&$14.773$&$0.213$&$-26.754$&$-$&$+$&$+$\\
$26$&CM2-W2&$-0.662$&$5.316$&$-0.904$&$0.213$&$18.090$&$0.213$&$-26.965$&$-$&$+$&$+$\\
$27$&CM2-W2&$-0.648$&$5.310$&$-2.109$&$0.453$&$13.204$&$0.608$&$-21.693$&$-$&$+$&$+$\\
$28$&CM2-W2&$-0.647$&$5.305$&$-10.133$&$0.168$&$14.082$&$0.168$&$-10.530$&$-$&$+$&$+$\\
$29$&CM2-W2&$-0.654$&$5.306$&$3.759$&$2.676$&$11.554$&$2.356$&$-27.802$&$-$&$+$&$+$\\
$30$&CM2-W2&$-0.654$&$5.306$&$-10.637$&$0.259$&$35.577$&$0.259$&$-28.434$&$-$&$+$&$+$\\
$31$&CM2-W2&$-0.657$&$5.306$&$-8.152$&$0.168$&$27.284$&$0.168$&$-26.330$&$-$&$+$&$+$\\
$32$&CM2-W2&$-0.690$&$5.297$&$-4.221$&$0.213$&$11.909$&$0.213$&$-5.055$&$-$&$+$&$+$\\
$33$&CM2-W2&$-0.690$&$5.297$&$-0.350$&$0.168$&$6.364$&$0.168$&$-5.685$&$-$&$+$&$+$\\
$34$&CM2-W2&$-0.691$&$5.298$&$9.949$&$0.213$&$-0.603$&$0.213$&$-6.741$&$-$&$+$&$+$\\
$35$&CM2-W2&$-0.691$&$5.298$&$-1.809$&$0.213$&$12.211$&$0.213$&$-7.162$&$-$&$+$&$+$\\
$36$&CM2-W2&$-0.691$&$5.298$&$-1.507$&$0.213$&$11.758$&$0.213$&$-7.373$&$-$&$+$&$+$\\
$37$&CM2-W2&$-0.690$&$5.298$&$-3.301$&$0.259$&$7.886$&$0.259$&$-9.896$&$-$&$+$&$+$\\
$38$&CM2-W2&$-0.686$&$5.304$&$-4.629$&$1.477$&$17.235$&$0.324$&$-11.373$&$-$&$+$&$+$\\
$39$&CM2-W2&$-0.685$&$5.304$&$1.809$&$0.213$&$12.814$&$0.213$&$-13.693$&$-$&$+$&$+$\\
$40$&CM2-W2&$-0.685$&$5.304$&$1.206$&$0.213$&$12.060$&$0.213$&$-13.903$&$-$&$+$&$+$\\
$41$&CM2-W2&$-0.637$&$5.367$&$-1.583$&$0.306$&$16.884$&$0.594$&$-14.746$&$-$&$+$&$+$\\
$42$&CM2-W2&$-0.642$&$5.363$&$4.899$&$1.253$&$26.306$&$0.841$&$-13.693$&$-$&$+$&$+$\\
$43$&CM2-W2&$-0.643$&$5.365$&$-3.668$&$0.205$&$22.307$&$0.217$&$-14.322$&$-$&$+$&$+$\\
$44$&CM2-W2&$-0.640$&$5.361$&$-6.683$&$1.455$&$20.150$&$2.021$&$-10.954$&$-$&$+$&$+$\\
$45$&CM2-W2&$-0.627$&$5.379$&$0.377$&$0.642$&$19.823$&$0.306$&$0.633$&$-$&$+$&$+$\\
$46$&CM2-W2&$-0.628$&$5.378$&$2.110$&$0.213$&$17.336$&$0.213$&$-5.055$&$-$&$+$&$+$\\
$47$&CM2-W2&$-0.647$&$5.340$&$-1.324$&$0.117$&$15.224$&$0.117$&$-5.474$&$-$&$+$&$+$\\
$48$&CM2-W2&$-0.660$&$5.353$&$-0.937$&$0.168$&$11.078$&$0.168$&$-3.789$&$-$&$+$&$+$\\
$49$&CM2-W2&$-0.486$&$5.330$&$4.372$&$0.213$&$15.678$&$0.213$&$0.633$&$+$&$+$&$+$\\
$50$&CM2-W2&$-0.486$&$5.330$&$4.975$&$0.213$&$12.663$&$0.213$&$0.422$&$+$&$+$&$+$\\
$51$&CM2-W2&$-0.486$&$5.330$&$5.395$&$0.168$&$10.848$&$0.168$&$-0.418$&$+$&$+$&$+$\\
$52$&CM2-W2&$-0.486$&$5.330$&$0.627$&$0.168$&$21.485$&$0.168$&$-0.629$&$+$&$+$&$+$\\
$53$&CM2-W2&$-0.486$&$5.330$&$5.379$&$0.168$&$9.349$&$0.168$&$-0.840$&$+$&$+$&$+$\\
$54$&CM2-W2&$-0.486$&$5.330$&$6.037$&$0.168$&$11.323$&$0.168$&$-1.050$&$+$&$+$&$+$\\
$55$&CM2-W2&$-0.486$&$5.330$&$5.611$&$0.168$&$8.848$&$0.168$&$-1.261$&$+$&$+$&$+$\\
$56$&CM2-W2&$-0.486$&$5.330$&$-2.550$&$0.168$&$19.441$&$0.168$&$-1.893$&$+$&$+$&$+$\\
$57$&CM2-W2&$-0.572$&$5.405$&$0.452$&$0.213$&$24.120$&$0.213$&$-45.082$&$+$&$+$&$+$\\
$58$&CM2-W2&$-0.572$&$5.405$&$-0.904$&$0.213$&$22.612$&$0.213$&$-45.293$&$+$&$+$&$+$\\
$59$&CM2-W2&$-0.572$&$5.405$&$3.698$&$0.168$&$24.442$&$0.168$&$-46.975$&$+$&$+$&$+$\\
$60$&CM2-W2&$-0.602$&$5.378$&$0.301$&$0.433$&$17.487$&$0.406$&$-21.909$&$+$&$+$&$+$\\
$61$&CM2-W2&$-0.690$&$5.289$&$-1.585$&$0.168$&$16.134$&$0.168$&$-6.528$&$+$&$+$&$+$\\
$62$&CM2-W2&$-0.662$&$5.317$&$-4.279$&$0.168$&$28.030$&$0.168$&$-21.485$&$+$&$+$&$+$\\
$63$&CM2-W2&$-0.656$&$5.307$&$2.384$&$0.259$&$8.436$&$0.259$&$-26.959$&$+$&$+$&$+$\\
$64$&CM2-W2&$-0.659$&$5.308$&$-1.055$&$0.213$&$26.532$&$0.213$&$-26.754$&$+$&$+$&$+$\\
$65$&CM2-W2&$-0.659$&$5.309$&$4.768$&$0.259$&$20.723$&$0.259$&$-26.117$&$+$&$+$&$+$\\
$66$&CM2-W2&$-0.614$&$5.347$&$1.507$&$0.213$&$23.215$&$0.213$&$-19.591$&$+$&$+$&$+$\\
$67$&CM2-W2&$-0.690$&$5.296$&$0.754$&$0.213$&$14.924$&$0.213$&$-4.844$&$+$&$+$&$+$\\
$68$&CM2-W2&$-0.688$&$5.304$&$3.618$&$0.213$&$5.125$&$0.213$&$-4.634$&$+$&$+$&$+$\\
$69$&CM2-W2&$-0.688$&$5.304$&$1.131$&$2.624$&$13.718$&$0.337$&$-5.266$&$+$&$+$&$+$\\
$70$&CM2-W2&$-0.688$&$5.305$&$3.481$&$0.168$&$15.810$&$0.168$&$-7.160$&$+$&$+$&$+$\\
$71$&CM2-W2&$-0.688$&$5.305$&$-1.013$&$0.168$&$15.293$&$0.168$&$-7.370$&$+$&$+$&$+$\\
$72$&CM2-W2&$-0.688$&$5.305$&$-0.603$&$0.213$&$16.281$&$0.213$&$-7.583$&$+$&$+$&$+$\\
$73$&CM2-W2&$-0.686$&$5.304$&$-3.769$&$0.213$&$19.899$&$0.213$&$-8.637$&$+$&$+$&$+$\\
$74$&CM2-W2&$-0.686$&$5.305$&$5.578$&$0.213$&$5.728$&$0.213$&$-10.954$&$+$&$+$&$+$\\
$75$&CM2-W2&$-0.686$&$5.304$&$-3.951$&$0.168$&$14.085$&$0.168$&$-11.162$&$+$&$+$&$+$\\
$76$&CM2-W2&$-0.666$&$5.336$&$0.754$&$0.213$&$18.090$&$0.213$&$-9.058$&$+$&$+$&$+$\\
$77$&CM2-W2&$-0.682$&$5.327$&$1.357$&$0.213$&$20.954$&$0.213$&$-1.895$&$+$&$+$&$+$\\
$78$&CM2-W2&$-0.631$&$5.372$&$-0.357$&$0.251$&$23.737$&$0.282$&$-16.429$&$+$&$+$&$+$\\
$79$&CM2-W2&$-0.614$&$5.376$&$-1.009$&$2.068$&$21.915$&$0.701$&$-5.472$&$+$&$+$&$+$\\
$80$&CM2-W2&$-0.647$&$5.387$&$-5.578$&$0.282$&$17.185$&$0.337$&$-13.693$&$+$&$+$&$+$\\
$81$&CM2-W2&$-0.651$&$5.324$&$-3.392$&$0.117$&$38.307$&$0.117$&$-10.109$&$+$&$+$&$+$\\
$82$&CM2-W2&$-0.646$&$5.325$&$1.407$&$0.117$&$13.155$&$0.117$&$-23.802$&$+$&$+$&$+$\\
$83$&CM2-W2&$-0.646$&$5.326$&$4.035$&$0.259$&$15.955$&$0.259$&$-22.536$&$+$&$+$&$+$\\
$84$&CM2-W2&$-0.667$&$5.346$&$2.648$&$0.117$&$13.238$&$0.117$&$-3.368$&$+$&$+$&$+$\\
$85$&CM2-W2&$-0.668$&$5.346$&$0.183$&$0.259$&$9.536$&$0.259$&$-5.893$&$+$&$+$&$+$\\
$86$&CM2-W2&$-0.663$&$5.358$&$-1.658$&$0.213$&$12.814$&$0.213$&$-10.954$&$+$&$+$&$+$\\
$87$&CM2-W2&$-0.661$&$5.354$&$0.183$&$0.259$&$10.086$&$0.259$&$-2.944$&$+$&$+$&$+$\\
$88$&MM1-W1&$-0.222$&$3.641$&$-0.904$&$0.213$&$12.361$&$0.213$&$-14.325$&$+$&$+$&$-$\\
$89$&MM1-W1&$-0.223$&$3.641$&$-2.237$&$0.168$&$15.733$&$0.168$&$-13.901$&$+$&$+$&$-$\\
$90$&MM1-W1&$-0.222$&$3.641$&$-3.467$&$0.213$&$5.125$&$0.213$&$-14.535$&$+$&$+$&$-$\\
$91$&MM1-W1&$-0.222$&$3.641$&$-0.248$&$0.117$&$7.115$&$0.117$&$-14.111$&$+$&$+$&$-$\\
$92$&MM1-W1&$-0.188$&$3.672$&$-7.436$&$0.678$&$12.387$&$1.270$&$-61.497$&$+$&$+$&$-$\\
$93$&MM1-W1&$-0.188$&$3.672$&$-8.178$&$0.410$&$12.776$&$0.984$&$-63.094$&$+$&$+$&$-$\\
$94$&MM1-W1&$-0.188$&$3.674$&$-8.953$&$0.168$&$19.211$&$0.168$&$-64.039$&$+$&$+$&$-$\\
$95$&MM1-W1&$-0.223$&$3.641$&$-0.151$&$0.213$&$16.130$&$0.213$&$-13.693$&$+$&$+$&$+$\\
$96$&MM1-W1&$-0.188$&$3.674$&$-11.940$&$0.168$&$21.226$&$0.168$&$-63.407$&$+$&$+$&$+$\\
$97$&MM1-W1&$-0.188$&$3.674$&$-11.740$&$0.168$&$20.267$&$0.168$&$-62.986$&$+$&$+$&$+$\\
$98$&MM1-W1&$-0.188$&$3.674$&$-11.389$&$0.168$&$19.727$&$0.168$&$-62.564$&$+$&$+$&$+$\\
$99$&MM1-W1&$-0.188$&$3.674$&$-6.978$&$0.168$&$7.344$&$0.168$&$-61.511$&$+$&$+$&$+$\\
$100$&MM1-W1&$-0.188$&$3.674$&$-7.115$&$0.117$&$14.644$&$0.117$&$-60.458$&$+$&$+$&$+$\\
$101$&MM1-W3&$-0.272$&$2.884$&$0.301$&$0.213$&$2.110$&$0.213$&$-0.420$&$+$&$+$&$-$\\
$102$&MM1-W3&$-0.272$&$2.883$&$3.316$&$0.213$&$-0.754$&$0.213$&$-0.420$&$+$&$+$&$-$\\
$103$&MM1-W3&$-0.272$&$2.883$&$1.357$&$0.213$&$-1.507$&$0.213$&$-0.210$&$+$&$+$&$-$\\
$104$&MM1-W3&$-0.273$&$2.889$&$3.095$&$0.168$&$6.649$&$0.168$&$-3.789$&$+$&$+$&$-$\\
$105$&MM1-W3&$-0.273$&$2.892$&$-1.241$&$0.117$&$-1.241$&$0.117$&$-3.368$&$+$&$+$&$-$\\
$106$&MM1-W3&$-0.273$&$2.892$&$-3.806$&$0.117$&$-1.655$&$0.117$&$-3.789$&$+$&$+$&$-$\\
$107$&MM1-W3&$-0.274$&$2.892$&$-2.292$&$1.652$&$-9.261$&$2.806$&$-2.523$&$+$&$+$&$-$\\
$108$&MM1-W3&$-0.274$&$2.894$&$0.579$&$0.117$&$7.198$&$0.117$&$-1.472$&$+$&$+$&$-$\\
$109$&MM1-W3&$-0.274$&$2.894$&$0.678$&$0.168$&$8.936$&$0.168$&$-1.682$&$+$&$+$&$-$\\
$110$&MM1-W3&$-0.274$&$2.893$&$0.619$&$0.168$&$10.558$&$0.168$&$-1.893$&$+$&$+$&$-$\\
$111$&MM1-W3&$-0.274$&$2.893$&$-0.956$&$0.168$&$15.018$&$0.168$&$-2.314$&$+$&$+$&$-$\\
$112$&MM1-W3&$-0.274$&$2.897$&$2.952$&$0.359$&$11.163$&$2.192$&$-2.314$&$+$&$+$&$-$\\
$113$&MM1-W3&$-0.273$&$2.888$&$-1.055$&$0.213$&$11.758$&$0.213$&$-2.738$&$+$&$+$&$+$\\
$114$&MM1-W3&$-0.273$&$2.888$&$-3.937$&$0.213$&$9.222$&$1.266$&$-3.157$&$+$&$+$&$+$\\
$115$&MM1-W3&$-0.273$&$2.887$&$-0.904$&$0.213$&$17.487$&$0.213$&$-3.370$&$+$&$+$&$+$\\
$116$&MM1-W3&$-0.274$&$2.891$&$3.316$&$0.213$&$4.070$&$0.213$&$-2.316$&$+$&$+$&$+$\\
$117$&MM1-W3&$-0.274$&$2.891$&$0.083$&$0.117$&$-0.993$&$0.117$&$-2.736$&$+$&$+$&$+$\\
$118$&MM1-W3&$-0.273$&$2.890$&$-3.061$&$0.117$&$-1.986$&$0.117$&$-3.157$&$+$&$+$&$+$\\
$119$&MM1-W3&$-0.273$&$2.890$&$2.076$&$0.168$&$6.296$&$0.168$&$-3.368$&$+$&$+$&$+$\\
$120$&MM1-W3&$-0.273$&$2.890$&$2.319$&$0.168$&$6.502$&$0.168$&$-3.578$&$+$&$+$&$+$\\
$121$&MM1-W3&$-0.273$&$2.892$&$-2.068$&$0.117$&$-2.317$&$0.117$&$-3.578$&$+$&$+$&$+$\\
$122$&MM1-W3&$-0.274$&$2.891$&$-9.720$&$0.259$&$0.550$&$0.259$&$-2.312$&$+$&$+$&$+$\\
$123$&MM1-W3&$-0.274$&$2.893$&$1.661$&$0.168$&$4.084$&$0.168$&$-2.736$&$+$&$+$&$+$\\
$124$&MM1-W3&$-0.274$&$2.894$&$4.824$&$0.213$&$1.658$&$0.213$&$-3.159$&$+$&$+$&$+$\\
$125$&MM1-W3&$-0.274$&$2.893$&$-0.301$&$0.213$&$11.457$&$0.213$&$-3.370$&$+$&$+$&$+$\\
$126$&UCHII-W3&$0.137$&$1.277$&$2.834$&$0.168$&$-10.437$&$0.168$&$-36.231$&$-$&$+$&$+$\\
$127$&UCHII-W3&$0.137$&$1.277$&$4.070$&$0.213$&$-12.512$&$0.213$&$-36.655$&$-$&$+$&$+$\\
$128$&UCHII-W3&$0.178$&$1.274$&$4.774$&$1.483$&$-8.751$&$0.393$&$-47.607$&$+$&$+$&$+$\\
$129$&UCHII-W3&$0.178$&$1.274$&$6.555$&$0.168$&$-16.538$&$0.168$&$-48.028$&$+$&$+$&$+$\\
$130$&UCHII-W1&$-0.072$&$-0.060$&$0.301$&$0.213$&$8.743$&$0.213$&$-14.957$&$+$&$+$&$-$\\
$131$&UCHII-W1&$-0.072$&$-0.060$&$2.044$&$0.168$&$4.421$&$0.168$&$-15.165$&$+$&$+$&$-$\\
$132$&UCHII-W1&$-0.072$&$-0.064$&$-1.960$&$0.500$&$10.251$&$1.830$&$-13.482$&$-$&$+$&$+$\\
$133$&UCHII-W1&$-0.072$&$-0.063$&$-4.673$&$0.213$&$7.537$&$0.213$&$-13.693$&$-$&$+$&$+$\\
$134$&UCHII-W1&$-0.072$&$-0.064$&$-3.166$&$0.994$&$12.889$&$1.778$&$-13.903$&$-$&$+$&$+$\\
$135$&UCHII-W1&$-0.072$&$-0.060$&$0.151$&$0.213$&$9.346$&$0.213$&$-15.378$&$-$&$+$&$+$\\
$136$&UCHII-W1&$-0.074$&$-0.059$&$-0.603$&$0.213$&$9.196$&$0.213$&$-13.271$&$-$&$+$&$+$\\
$137$&UCHII-W1&$-0.074$&$-0.059$&$-0.754$&$0.213$&$9.497$&$0.213$&$-13.482$&$-$&$+$&$+$\\
$138$&UCHII-W1&$-0.074$&$-0.059$&$0.092$&$0.168$&$7.869$&$0.168$&$-14.322$&$-$&$+$&$+$\\
$139$&UCHII-W1&$-0.074$&$-0.059$&$-0.285$&$0.168$&$8.164$&$0.168$&$-14.533$&$-$&$+$&$+$\\
$140$&UCHII-W1&$-0.074$&$-0.058$&$-0.301$&$0.213$&$3.618$&$0.213$&$-14.746$&$-$&$+$&$+$\\
$141$&UCHII-W1&$-0.074$&$-0.059$&$-6.602$&$0.259$&$20.173$&$0.259$&$-12.635$&$-$&$+$&$+$\\
$142$&UCHII-W1&$0.001$&$-0.000$&$1.737$&$0.117$&$-1.489$&$0.117$&$-19.167$&$+$&$+$&$+$\\
$143$&UCHII-W1&$0.001$&$-0.001$&$-2.640$&$0.168$&$1.697$&$0.168$&$-19.589$&$+$&$+$&$+$\\
$144$&UCHII-W1&$0.002$&$-0.001$&$-0.301$&$0.213$&$-1.357$&$0.213$&$-22.119$&$+$&$+$&$+$\\
$145$&UCHII-W1&$0.001$&$-0.001$&$0.301$&$0.213$&$0.301$&$0.213$&$-19.381$&$+$&$+$&$+$\\
$146$&UCHII-W1&$0.000$&$-0.000$&$---$&$---$&$---$&$---$&$-20.642$&$+$&$+$&$+$\\
$147$&UCHII-W1&$0.000$&$-0.000$&$---$&$---$&$---$&$---$&$-21.063$&$+$&$+$&$+$\\
$148$&UCHII-W1&$0.000$&$-0.000$&$---$&$---$&$---$&$---$&$-21.274$&$+$&$+$&$+$\\
$149$&UCHII-W1&$0.000$&$-0.000$&$---$&$---$&$---$&$---$&$-21.485$&$+$&$+$&$+$\\
$150$&UCHII-W1&$0.000$&$-0.000$&$---$&$---$&$---$&$---$&$-21.698$&$+$&$+$&$+$\\
$151$&UCHII-W1&$-0.072$&$-0.063$&$-0.754$&$0.213$&$9.196$&$0.213$&$-11.165$&$+$&$+$&$+$\\
$152$&UCHII-W1&$-0.072$&$-0.064$&$-1.658$&$0.213$&$11.608$&$0.213$&$-12.007$&$+$&$+$&$+$\\
$153$&UCHII-W1&$-0.072$&$-0.063$&$-2.261$&$0.213$&$11.457$&$0.213$&$-14.114$&$+$&$+$&$+$\\
$154$&UCHII-W1&$-0.072$&$-0.063$&$-2.858$&$0.201$&$7.751$&$1.055$&$-14.322$&$+$&$+$&$+$\\
$155$&UCHII-W1&$-0.072$&$-0.063$&$-7.702$&$0.259$&$-3.118$&$0.259$&$-12.635$&$+$&$+$&$+$\\
$156$&UCHII-W1&$-0.072$&$-0.059$&$0.603$&$0.213$&$15.678$&$0.213$&$-13.061$&$+$&$+$&$+$\\
$157$&UCHII-W1&$-0.072$&$-0.058$&$2.713$&$0.213$&$20.049$&$0.213$&$-13.271$&$+$&$+$&$+$\\
$158$&UCHII-W1&$-0.072$&$-0.058$&$-2.713$&$0.213$&$5.125$&$0.213$&$-14.746$&$+$&$+$&$+$\\
$159$&UCHII-W1&$-0.072$&$-0.058$&$-2.813$&$0.117$&$-5.130$&$0.117$&$-15.165$&$+$&$+$&$+$\\
$160$&UCHII-W1&$-0.071$&$-0.062$&$-3.478$&$0.168$&$11.494$&$0.168$&$-8.213$&$+$&$+$&$+$\\
$161$&UCHII-W1&$-0.071$&$-0.062$&$-2.377$&$0.168$&$11.660$&$0.168$&$-8.424$&$+$&$+$&$+$\\
$162$&UCHII-W1&$-0.072$&$-0.061$&$0.236$&$0.168$&$8.838$&$0.168$&$-9.477$&$+$&$+$&$+$\\
$163$&UCHII-W1&$-0.072$&$-0.061$&$0.026$&$0.168$&$10.197$&$0.168$&$-9.898$&$+$&$+$&$+$\\
$164$&UCHII-W1&$-0.072$&$-0.061$&$-1.950$&$0.168$&$12.685$&$0.168$&$-10.109$&$+$&$+$&$+$\\
$165$&UCHII-W1&$-0.072$&$-0.062$&$-2.262$&$0.168$&$18.775$&$0.168$&$-10.320$&$+$&$+$&$+$\\
$166$&UCHII-W1&$-0.072$&$-0.062$&$-1.960$&$0.213$&$22.461$&$0.213$&$-10.743$&$+$&$+$&$+$\\
$167$&UCHII-W1&$-0.072$&$-0.060$&$-5.125$&$0.213$&$-7.537$&$0.213$&$-10.954$&$+$&$+$&$+$\\
$168$&UCHII-W1&$-0.072$&$-0.061$&$-0.904$&$0.213$&$11.608$&$0.213$&$-11.165$&$+$&$+$&$+$\\
$169$&UCHII-W1&$-0.072$&$-0.061$&$1.498$&$0.168$&$12.758$&$0.168$&$-11.373$&$+$&$+$&$+$\\
$170$&UCHII-W1&$-0.072$&$-0.061$&$2.180$&$0.168$&$3.495$&$0.168$&$-11.584$&$+$&$+$&$+$\\
$171$&UCHII-W1&$-0.072$&$-0.061$&$1.159$&$0.168$&$4.015$&$0.168$&$-11.794$&$+$&$+$&$+$\\
$172$&UCHII-W1&$-0.072$&$-0.061$&$-2.718$&$0.168$&$14.095$&$0.168$&$-12.005$&$+$&$+$&$+$\\
$173$&UCHII-W1&$-0.072$&$-0.061$&$0.504$&$0.168$&$5.791$&$0.168$&$-12.216$&$+$&$+$&$+$\\
$174$&UCHII-W1&$-0.072$&$-0.061$&$0.330$&$0.168$&$6.882$&$0.168$&$-12.426$&$+$&$+$&$+$\\
$175$&UCHII-W1&$-0.072$&$-0.061$&$0.072$&$0.168$&$7.216$&$0.168$&$-12.637$&$+$&$+$&$+$\\
$176$&UCHII-W1&$-0.072$&$-0.061$&$-0.204$&$0.168$&$12.544$&$0.168$&$-13.690$&$+$&$+$&$+$\\
$177$&UCHII-W1&$-0.072$&$-0.061$&$-0.145$&$0.168$&$11.601$&$0.168$&$-13.901$&$+$&$+$&$+$\\
$178$&UCHII-W1&$-0.072$&$-0.061$&$0.410$&$0.168$&$9.786$&$0.168$&$-14.322$&$+$&$+$&$+$\\
$179$&UCHII-W1&$-0.072$&$-0.060$&$0.210$&$0.168$&$9.500$&$0.168$&$-14.533$&$+$&$+$&$+$\\
$180$&UCHII-W1&$-0.072$&$-0.060$&$2.211$&$0.168$&$2.420$&$0.168$&$-14.744$&$+$&$+$&$+$\\
$181$&UCHII-W1&$-0.072$&$-0.060$&$0.603$&$0.213$&$10.100$&$0.213$&$-15.799$&$+$&$+$&$+$\\
$182$&UCHII-W1&$-0.072$&$-0.060$&$0.868$&$0.168$&$2.213$&$0.168$&$-16.007$&$+$&$+$&$+$\\
$183$&UCHII-W1&$-0.072$&$-0.060$&$-8.619$&$0.259$&$22.557$&$0.259$&$-12.845$&$+$&$+$&$+$\\
$184$&UCHII-W1&$-0.074$&$-0.059$&$-0.301$&$0.213$&$10.402$&$0.213$&$-13.061$&$+$&$+$&$+$\\
$185$&UCHII-W1&$-0.074$&$-0.059$&$1.500$&$0.168$&$9.307$&$0.168$&$-13.901$&$+$&$+$&$+$\\
$186$&UCHII-W1&$-0.074$&$-0.059$&$0.507$&$0.168$&$7.296$&$0.168$&$-14.112$&$+$&$+$&$+$\\
 \enddata
\label{tab:pm_data}
\tablecomments{$^\text{a}$ Maser feature ID. \\
$^\text{b}$Offsets are with respect to the reference maser at $(\alpha,\delta)$ = (17$^\text{h}$20'52.600", -35$^\circ$46'50.508")}
\end{deluxetable*}

\begin{deluxetable*}{cccccccc}
\tablecaption{Position and Velocity Variance/Covariance Matrix Analysis for the \ngc\/ proper motions}
\tablewidth{\textwidth}
\tablehead{
\multicolumn{8}{c}{Diagonalization of the Position Variance/Covariance Matrices} \\
\cline{1-8}
Matrix No.$^\text{a}$ & $\psi_\text{max}$ & & $\psi_\text{min}$ & & PA$_\text{max}^b$ & &  \\
& (10$^{-6}$arcsec$^2$) & & (10$^{-6}$arcsec$^2$) & & ($^\circ$) & & 
}
\startdata
1 & 0.897 & & 0.003 & & 79.3  \\
2 & 0.545 & & 0.002 & & 79.0  & & \\
3 & 0.152 & & 0.0001 & & -84.4  & & \\
\cline{1-8}
\multicolumn{8}{c}{Diagonalization of the Velocity Variance/Covariance Matrices} \\
\cline{1-8}
Matrix No. & $\psi_\text{max}$ & $\psi_\text{mid}$ & $\psi_\text{min}$ & PA$_\text{max}$& PA$_\text{mid}^c$ & $\phi_\text{max}^d$ & $\phi_\text{mid}^e$ \\
\cline{1-8}
& (km$^2$s$^{-2}$) & (km$^2$s$^{-2}$) & (km$^2$s$^{-2}$) & ($^\circ$) & ($^\circ$) & ($^\circ$) & ($^\circ$) \\
\cline{1-8}
1 & 3511.08 & 452.48 & 201.79 & -79.4 $\pm$ 9.2 & 8.8 $\pm$ 1.1 & -32.4 $\pm$ 3.2 & -2.9 $\pm$ 0.3 \\
2 & 3321.62 & 274.64 & 148.94 & -77.9 $\pm$ 10.6 & 11.8 $\pm$ 1.5 & -6.0 $\pm$ 0.6 & -2.9 $\pm$ 0.3 \\
3 & 2604.80 & 582.89 & 93.42 & -67.1 $\pm$ 12.7 & 63.5 $\pm$ 12.0 & 42.8 $\pm$ 7.6 & 35.1 $\pm$ 6.3
\enddata
\tablecomments{\\ $^\text{a}$1: CM2-W2, MM1-W1, MM1-W3 \& UCHII-W3. All the maser regions associated with the jet. \\
2: CM2-W2 \& UCHII-W3. The maser regions furthest from MM1B \\
3: MM1-W1 \& MM1-W3. The maser regions closest to MM1B \\
$^\text{b}$ Position angle of the axis with the largest eigenvalue $\psi_\text{max}$ \\
$^\text{c}$ Position angle of the axis with the second largest eigenvalue $\psi_\text{mid}$ \\
$^\text{d}$ Inclination angle of the axis corresponding to $\psi_\text{max}$ with respect to the sky plane. \\
$^\text{e}$ Inclination angle of the axis corresponding to $\psi_\text{mid}$ with respect to the sky plane. }
\label{tab:PVCM/VVCM_results}
\end{deluxetable*}




\end{document}